\documentclass[%
 aip,
 jmp,%
 amsmath,amssymb,
 reprint,%
]{revtex4-1}

\usepackage{graphicx}
\usepackage{dcolumn}
\def\be{\begin{equation}}
\def\ee{\end{equation}}
\def\bea{\begin{array}}
\def\eea{\end{array}}
\def\beqa{\begin{eqnarray}}
\def\eeqa{\end{eqnarray}}
\def\beqas{\begin{eqnarray*}}
\def\eeqas{\end{eqnarray*}}

\def\bp{\begin{picture}}
\def\ep{\end{picture}}
\def\bc{\begin{center}}
\def\ec{\end{center}}
\def\bfig{\begin{figure}}
\def\efig{\end{figure}}

\def\bit{\begin{itemize}}
\def\eit{\end{itemize}}
\def\nn{\nonumber}
\def\f{\frac}

\def\[{\left[}
\def\]{\right]}
\def\({\left(}
\def\){\right)}

\def\..{\left.}
\def\.{\right.}
\def\tl{\tilde}

\def\la{\leftarrow}

\def\tm{\times}

\def\la{\lambda}

\def\al{\alpha}

\def\ep{\epsilon}
\def\rh{\rho}

\begin{document}

\preprint{AIP/123-QED}

\title{Generalized Froggatt-Nielsen Mechanism}

\author{Fei Wang}
\email{fei.wang@sci.monash.edu.au}
\affiliation{
School of Physics, Monash University, Melbourne Victoria 3800, Australia}%

\date{\today}

\begin{abstract}
In this paper, we propose a Generalized Froggatt-Nielsen mechanism in which non-renormalizable operators involving a GUT group and $U(1)_H$ non-singlet Higgs field are introduced. Thus the GUT gauge symmetry breaking and the generation of hierarchical flavor hierarchy have a common origin in this mechanism.  In this Generalized Froggatt-Nielsen mechanism, we propose universality conditions for coefficients corresponding to different contractions in the group productions. We find that the predictions in Generalized Froggatt-Nielsen mechanism for SU(5) GUT is different to that of ordinary Froggatt-Nielsen mechanism. Such Generalized Froggatt-Nielsen mechanism can be used in GUT models when ordinary Froggatt-Nielsen mechanism is no longer available. We study the application of Generalized Froggatt-Nielsen mechanism in SO(10) model. We find that realistic standard model mass hierarchy and mixings can be obtained both in SU(5) and SO(10) GUT models with such Generalized Froggatt-Nielsen mechanism. \end{abstract}

\pacs{Valid PACS appear here}
\maketitle

\section{\label{sec-0}Introduction}
 The standard model (SM) of electroweak interactions, based on the spontaneously broken
$SU(2)_L\tm U(1)_Y$ gauge symmetry, has been extremely successful in describing phenomena
below the weak scale. One of the longstanding puzzles of the standard model is the distinct pattern of masses and
mixing for the quarks and leptons which appear to be hierarchical in different generations. Froggatt-Nielsen mechanism\cite{FN} is an elegant solution to such puzzles in which standard model
renormalizable yukawa couplings are the low energy effective interactions of certain high energy non-renormalizable interactions with  $U(1)_H$ horizontal symmetry between different generations.

  The approximately unification of the
$SU(3)_C\times SU(2)_L \times U(1)_Y$ gauge couplings strongly suggests the existence of a Grand Unified Theory (GUT).
Grand Unification Theory, such as the $SU(5)$~\cite{Georgi:1974sy} and $SO(10)$~\cite{so10} models, give us deep insights into the
problems of the standard model such as charge quantization, the origin of many free parameters,
the standard model fermion masses and mixings, and beyond. In order to interpret the hierarchical standard model flavor structure, it is possible to use Froggatt-Nielsen mechanism in GUT models which is very predictive due to the unification of the standard model matter contents. However, the applications of ordinary Froggatt-Nielsen mechanism in GUT models are limited because it is very hard to get the realistic standard model hierarchy structure in many GUT models.

In ordinary Froggatt-Nielsen mechanism, renormalizable standard model yukawa couplings are not invariant under horizontal $U(1)_H$ gauge symmetry. One introduce the standard model gauge singlet Higgs field with non-trivial $U(1)_H$ quantum number to break the $U(1)_H$ horizontal symmetry which, after acquiring VEVs, can give rise to hierarchical standard model yukawa couplings from the high energy non-renormalizable $U(1)_H$ invariant Lagrangian. In GUT models, there are alternative possibilities in which we use certain high dimensional representation Higgs fields to break the GUT group as well as the $U(1)_H$ horizontal symmetry. Standard Model yukawa couplings can again be regarded as the low energy effective interactions of some high energy $U(1)_H$ invariant non-renormalizable interactions. This approach provides an unified point of view for both the GUT symmetry breaking and the generating of the flavor structure. We regard this approach as the Generalized Froggatt-Nielsen mechanism.

This paper is organized as follows. In sec-\ref{sec-1}, we discuss the Generalized Froggatt-Nielsen mechanism and its applications in SU(5) GUT models. In sec-\ref{sec-2}, we study the application of Generalized Froggatt-Nielsen mechanism in 4D SO(10) model and the generations of realistic standard model mass hierarchy with the simplest form for yukawa coupling. Sec-\ref{sec-3} contains our discussions and conclusions.

\section{\label{sec-1}Generalize Froggatt-Nielsen Mechanism In SU(5) Model}
 It is well known that the SM fermion masses and mixings exhibit a
hierarchical structure. The quark CKM mixings can be cast in the
Wolfenstein formalism as~\cite{CKM} \beqa
V_{CKM}=\(\bea{ccc}1-\f{\tl{\la}^2}{2}&\tl{\la}& A\tl{\la}^3(\rh+i\eta)\\
-\tl{\la}&1-\f{\tl{\la}^2}{2}&A\tl{\la}^2\\A\tl{\la}^3(1-\rh+i\eta)&~-A\tl{\la}^2&1\eea\),\nn\\
\eeqa where $A$ is of order 1 while $\rh$ and $\eta$ are between
$\tl{\la}$ and 1. The hierarchy is reflected in the dependence of various
entries on different powers of $\tl{\la}\approx 0.22$. Renormalization
group equation (RGE) running of the charged fermion masses to a high scale
($\sim 10^{16}$ GeV) also reveals the following hierarchical
structure \beqa\label{ratio} m_t~:~m_c~:~m_u~&\simeq&
~1~:~\tl{\la}^4~:~\tl{\la}^8~,\nn\\~m_b~:~m_s~:~m_d~&\simeq&
~1~:~\tl{\la}^2~:~\tl{\la}^4~,\nn\\~m_\tau~:~m_\mu~:~m_e~&\simeq&
~1~:~\tl{\la}^2~:~\tl{\la}^4~, \eeqa with $m_b/m_t=\tl{\la}^3$.
Grand Unification Theory which predicts the unification of matter contents should also explain the standard model hierarchical structure. So it is natural to think about the applications of Froggatt-Nielsen mechanism to GUT models. Ordinary Froggatt-Nielsen mechanism can be used in SU(5) GUT to get the desired flavor structure.

 In ordinary Froggatt-Nielsen mechanism, the standard model gauge singlet Higgs field with non-trivial $U(1)_H$ charge is introduced. In SU(5) GUT model, the $U(1)_H$ gauge invariant non-renormalizable Lagrangian is introduced
\beqa
{\cal L}&&\supset \al_{ij}\(\f{S}{M}\)^{(Q_{\bf 10}^i+Q_{\bf 10}^j+Q_{\bf H})} F_{\bf 10}^i\otimes F_{\bf 10}^j\otimes H_{\bf 5}\nn\\
&&+ \beta_{ij}\(\f{S}{M}\)^{(Q_{\bf 10}^i+Q_{\bf \bar{5}}^j+Q_{\bf \bar{H}})} F_{\bf 10}^i\otimes \bar{f}_{\bf \bar{5}}^j\otimes \bar{H}_{\bf \bar{5}},\nn\\
\eeqa
with some new physics energy scale $M$ and the $U(1)_H$ charge $Q(S)=-1$.
  After the GUT singlet Higgs field $S$ acquires a Vacuum Expectation Values $<S>$ to break the $U(1)_H$ horizontal gauge symmetry, the hierarchical standard model yukawa couplings are obtained.

For example, we can get the flavor structure by assigning the $U(1)_H$ charge
\beqa
&&Q(F_{\bf 10}^1)=4,~Q(F_{\bf 10}^2)=2~,Q(F_{\bf 10}^3)=0,\\
&&Q(\bar{f}_{\bf \bar{5}}^1)=0,~~Q(\bar{f}_{\bf \bar{5}}^2)=0,~~~Q(\bar{f}_{\bf \bar{5}}^3)=0,\\
&&Q(H)=0~,
\eeqa
to get the realistic mass ratio in (\ref{ratio})
with
\beqa
\tl{\la}=\f{<S>}{M}\approx 0.22~.
\eeqa
However, it is difficult to apply this ordinary Froggatt-Nielsen mechanism in other 4D GUT models. For example, we can not get realistic mass hierarchy in SO(10) GUT models with simplest yukawa coupling term when ordinary Froggatt-Nielsen mechanism is used. In this paper, we propose a Generalized Froggatt-Nielsen mechanism to solve such difficulties.

In our approach, the GUT group non-singlet Higgs fields $\Phi$ with non-trivial horizontal $U(1)_H$ charge is introduced. By assigning proper $U(1)_H$ quantum numbers to SU(5) matter contents $F_{\bf 10}^i$ and $\bar{f}_{\bf \bar{5}}^j$ , we can get the gauge invariant non-renormalizable Lagrangian
\beqa
{\cal L}&&\supset \sum\limits_s\al_{ij}^s\(\f{\Phi_{\bf 24}}{M}\)^{(Q_{\bf 10}^i+Q_{\bf 10}^j+Q_{\bf H})} F_{\bf 10}^i\otimes F_{\bf 10}^j\otimes H_{\bf 5}\nn\\
&&+\sum\limits_t\beta_{ij}^t\(\f{\Phi_{\bf 24}}{M}\)^{(Q_{\bf 10}^i+Q_{\bf \bar{5}}^j+Q_{\bf \bar{H}})} F_{\bf 10}^i\otimes \bar{f}_{\bf \bar{5}}^j\otimes \bar{H}_{\bf \bar{5}},
\eeqa
with the index $s$,$t$ denote the different group production contractions and the $U(1)_H$ charge $Q(\Phi)=-1$. The coefficients $\al_{ij}^s$ and $\beta_{ij}^t$ which corresponds to different contractions in the group production are ${\cal O}(1)$ parameters. The group production of such GUT group non-singlet Higgs fields will give various non-renormalizable operators, including terms that appear in ordinary Froggatt-Nielsen mechanism. As there are many types of contractions in the group production, the most naturally choice for the corresponding coefficients is by assuming their universality
\beqa
\al_{ij}^s=\al_{ij}~(\forall s);~~~~~~ \beta_{ij}^t=\beta_{ij}~ (\forall t).
\eeqa
The adjoint Higgs can acquire Vacuum Expectation Values (VEVs) to break the SU(5) GUT gauge group into $SU(3)_c\tm SU(2)_L\tm U(1)_Y$. Such VEVs can be written in terms of $5\tm 5$ matrix
\beqa
<\Phi_{\bf 24}>=v_0\sqrt{\f{3}{5}}{\rm diag}(~\f{1}{3},~\f{1}{3},~\f{1}{3},-\f{1}{2},-\f{1}{2}),
\eeqa
with normalization factor $c=\f{1}{2}$ as well as $10\tm 10$ matrix
\beqa
<\Phi_{\bf 24}>=v_0\sqrt{\f{3}{5}}{\rm diag}(\underbrace{\f{2}{3},\cdots, \f{2}{3}}_{3},\underbrace{-\f{1}{6},\cdots,-\f{1}{6}}_{6},-1),\nn\\
\eeqa
with normalization factor $c=\f{3}{2}$.

The predictions of generalized Froggatt-Nielsen mechanism in SU(5) GUT is different to that of ordinary Froggatt-Nielsen mechanism. The mass matrix for up-type quarks can be calculated to be
\beqa 
\tiny
\(\bea{ccc}
~1~&\(\f{2}{3}\la\)^{a_{12}}+\(-\f{1}{6}\la\)^{a_{12}}&\(\f{2}{3}\la\)^{a_{13}}+\(-\f{1}{6}\la\)^{a_{13}}\\
\(\f{2}{3}\la\)^{a_{12}}+\(-\f{1}{6}\la\)^{a_{12}}&\(\f{2}{3}\la\)^{a_{22}}+\(-\f{1}{6}\la\)^{a_{22}}&\(\f{2}{3}\la\)^{a_{23}}+\(-\f{1}{6}\la\)^{a_{23}}\\
\(\f{2}{3}\la\)^{a_{13}}+\(-\f{1}{6}\la\)^{a_{13}}&\(\f{2}{3}\la\)^{a_{23}}+\(-\f{1}{6}\la\)^{a_{23}}&\(\f{2}{3}\la\)^{a_{33}}+\(-\f{1}{6}\la\)^{a_{33}}
\eea\),\nn
\eeqa
which after simplification 
\beqa
{\cal M}_u&\approx&
\(\bea{ccc}
~1~~~&~\(\f{2}{3}\la\)^{(a_{12})}&~~~\(\f{2}{3}\la\)^{(a_{13})}\\
~~\(\f{2}{3}\la\)^{(a_{12})}&~~\(\f{2}{3}\la\)^{(a_{22})}&~~\(\f{2}{3}\la\)^{(a_{23})}\\
~\(\f{2}{3}\la\)^{(a_{13})}&~~\(\f{2}{3}\la\)^{(a_{23})}&~~\(\f{2}{3}\la\)^{(a_{33})}
\eea\),\nn\\
\eeqa
will give the mass hierarchy
\beqa
m_t:m_c:m_u\approx ~1:\(\f{2}{3}\la\)^{(a_{22})}:\(\f{2}{3}\la\)^{(a_{33})}.
\eeqa
Here we use the notation
\beqa
\la=\sqrt{\f{3}{5}}\f{v_0}{M},
\eeqa
and the definition
\beqa a_{ij}[U]{\equiv}Q[U^i_L]+Q[(U^c_L)^j]=Q[{\bf 10}^i]+Q[{\bf 10}^j]~.\eeqa

 We can chose
\beqa
Q[{\bf 10}^1]=2,~Q[{\bf 10}^2]=1~,Q[{\bf 10}^3]=0~,
\eeqa
with
\beqa
\f{2}{3}\la\approx (0.22)^2,
\eeqa
to get the realistic mass hierarchy for up-type quarks.

Similarly, we can caculate the mass matrix for down-type quarks
\beqa 
\tiny
\(\bea{ccc}
~1~&\(-\f{1}{6}\la\)^{b_{12}}-\(\f{1}{2}\la\)^{b_{12}}&\(-\f{1}{6}\la\)^{b_{13}}-\(\f{1}{2}\la\)^{b_{13}}\\
\(-\f{1}{6}\la\)^{b_{21}}-\(\f{1}{2}\la\)^{b_{21}}&\(-\f{1}{6}\la\)^{b_{22}}-\(\f{1}{2}\la\)^{b_{22}}&\(-\f{1}{6}\la\)^{b_{23}}-\(\f{1}{2}\la\)^{b_{23}}\\
\(-\f{1}{6}\la\)^{b_{31}}-\(\f{1}{2}\la\)^{b_{31}}&\(-\f{1}{6}\la\)^{b_{32}}-\(\f{1}{2}\la\)^{b_{32}}&\(-\f{1}{6}\la\)^{b_{33}}-\(\f{1}{2}\la\)^{b_{33}}\\
\eea\),\nn\eeqa
and charged leptons
\beqa 
\tiny
\(\bea{ccc}
~1~~~&~\(\f{1}{2}\la\)^{(b_{12})}+\la^{(b_{12})}&~~~\(\f{1}{2}\la\)^{(b_{13})}+\la^{(b_{13})}\\
\(\f{1}{2}\la\)^{(b_{21})}+\la^{(b_{21})}~~~&~\(\f{1}{2}\la\)^{(b_{22})}+\la^{(b_{22})}&~~~\(\f{1}{2}\la\)^{(b_{23})}+\la^{(b_{23})}\\
\(\f{1}{2}\la\)^{(b_{31})}+\la^{(b_{31})}~~~&~\(\f{1}{2}\la\)^{(b_{32})}+\la^{(b_{32})}&~~~\(\f{1}{2}\la\)^{(b_{33})}+\la^{(b_{33})}
\eea\),\nn
\eeqa
with the definition
\beqa
b_{ij}[D]=b_{ij}[E]=Q[{\bf 10}^i]+Q[{\bf \bar{5}}^j]~.
\eeqa
After simplification, they will give 
\beqa
{\cal M}_d&\approx&
\(\bea{ccc}
~1~~~&-\(\f{1}{2}\la\)^{b_{12}}&~-\(\f{1}{2}\la\)^{b_{13}}\\
-\(\f{1}{2}\la\)^{b_{21}}~&-\(\f{1}{2}\la\)^{b_{22}}&~-\(\f{1}{2}\la\)^{b_{23}}\\
-\(\f{1}{2}\la\)^{b_{31}}~&-\(\f{1}{2}\la\)^{b_{32}}&~-\(\f{1}{2}\la\)^{b_{33}}
\eea\),
\eeqa
and 
\beqa
{\cal M}_e\approx&
\(\bea{ccc}
~1~~~&~\la^{(b_{12})}&~~~\la^{(b_{13})}\\
~\la^{(b_{21})}~~&~\la^{(b_{22})}&~~~\la^{(b_{23})}\\
\la^{(b_{31})}~~&~\la^{(b_{32})}&~~~\la^{(b_{33})}\\
\eea\).
\eeqa
Then we can get approximately the realistic mass hierarchy for down-type quarks and charged leptons
\beqa
m_b:m_s:m_d&\approx& ~1:\(\f{1}{2}\la\)^{(b_{22})}:\(\f{1}{2}\la\)^{(b_{33})},\nn\\
m_\tau:m_\mu:m_e&\approx& ~1: \la^{(b_{22})}: \la^{(b_{33})}.
\eeqa
with the $U(1)_H$ charge assignment
\beqa
Q[{\bf \bar{5}}^1]=0,~Q[{\bf \bar{5}}^2]=0~,Q[{\bf \bar{5}}^3]=0~,
\eeqa
The mixing matrix can be easily obtained from the mass matrix and we will not give the expressions here.

  We can also use ${\bf 75},{\bf 200}, \cdots$ dimensional representation fields as $\Phi$ to construct the non-renormalizable terms. For example, the VEVs of ${\bf 75}$ dimensional representations can be written as
$10\tm 10$ matrix
\beqa
<\Phi_{\bf 75}>=\f{v_0}{2\sqrt{3}}{\rm diag}(\underbrace{~1,\cdots, ~1}_{6},\underbrace{-1,\cdots,-1}_{3},-3),
\eeqa
with normalization factor $c=\f{3}{2}$. Constructing Generalized Froggatt-Nielsen type non-normalizable terms invovling $\Phi_{\bf 75}$ and assuming universality for the coefficients corresponding to different contractions in the group production, we can obtain the mass hierarchy for standard model fermions
\beqa
m_t:m_c:m_u&\approx& 1: \la^{(a_{22})}: \la^{(a_{33})}~,\\
m_b:m_s:m_d&\approx& 1: \la^{(b_{12})}: \la^{(b_{13})}~,\\
m_\tau:m_\mu:m_e&\approx& 1: (3\la)^{(b_{12})}: (3\la)^{(b_{13})}~.
\eeqa
with
\beqa
\la=\f{v_0}{2\sqrt{3}M}\approx (0.22)^2.
\eeqa
Again, we can get approximately the realistic mass hierarchy by assigning the $U(1)_H$ charge
\beqa
&&Q[{\bf 10}^1]=2,~Q[{\bf 10}^2]=1~,Q[{\bf 10}^3]=0~,\nn\\
&&Q[{\bf \bar{5}}^1]=0,~Q[{\bf \bar{5}}^2]=0~,Q[{\bf \bar{5}}^3]=0~.
\eeqa

\section{\label{sec-2}Generalize Froggatt-Nielsen Mechanism In SO(10) Model}
SO(10) GUT model is very predictive not only because its genuine unification of all the standard model gauge couplings but also because all the standard model matter contents within each generation can be filled into one ${\bf 16}$ dimensional spinor representation. Due to its highly predictive nature (very few parameters), ordinary Froggatt-Nielsen mechanism can not be used in such 4D SO(10) GUT models to give the realistic standard model flavor structure. However, it is possible to use Generalized Froggatt-Nielsen mechanism in 4D SO(10) GUT models to obtain the realistic hierarchical flavor structure. We consider the following $U(1)_H$ invariant non-renormalizable Lagrangian
\beqa
{\cal L}\supset \sum\limits_{s}y_{ij}^s\(\f{\Phi_{\bf 45}}{M}\)^{(Q_{\bf 16}^i+Q_{\bf 16}^j+Q_{\bf H})}F_{\bf 16}^i F_{\bf 16}^j H_{\bf 10}~.
\eeqa
Again, the coefficients $y_{ij}^s$ which correspond to different contractions in the group production are ${\cal O}(1)$ parameters.  We also assume universality for each coefficients that corresponds to different contractions
\beqa
y_{ij}^s=y_{ij}~(\forall s) ~.
\eeqa

 The VEVs of {\bf 45} dimensional representation Higgs can break the SO(10) GUT model into Georgi-Glashow $SU(5)\tm U(1)$ model, flipped SU(5) model etc. As an example, we consider the symmetry breaking chain into Georgi-Glashow $SU(5)\tm U(1)$ model.
The VEVs of {\bf 45} dimensional representation Higgs can be written as $16\tm 16$ matrix
\beqa
<{\bf 45}>=\f{v_0}{6\sqrt{10}}(\underbrace{-1,\cdots,-1}_5,\underbrace{~\f{1}{3},\cdots,~\f{1}{3}}_{10},~\f{5}{3}),
\eeqa
with normalization factor $c=2$.
After calculations, we can obtain the prediction for the standard model mass hierarchy
\beqa
m_t:m_c:m_u&\approx& 1: (\f{1}{3}\la)^{(a_{22})}: (\f{1}{3}\la)^{(a_{33})}~,\\
m_b:m_s:m_d&\approx& 1: (\la)^{(a_{22})}: (\la)^{(a_{33})}~,\\
m_\tau:m_\mu:m_e&\approx& 1: (\la)^{(a_{22})}: (\la)^{(a_{33})}~.
\eeqa
with
\beqa
\la=\f{1}{6\sqrt{10}}\f{v_0}{M},
\eeqa
and
\beqa
a_{ij}=Q[{\bf 16}^i]+Q[{\bf 16}^j].
\eeqa
 We can get approximately the realistic mass hierarchy of standard model fermions by the $U(1)_H$ assignment
 \beqa
 Q[{\bf 16}^1]=3,~ Q[{\bf 16}^2]=\f{3}{2},~Q[{\bf 16}^3]=0,
 \eeqa
with the choice
\beqa
\la\approx (0.22)^{\f{2}{3}}~.
\eeqa
We can see that realistic standard model mass and mixing hierarchy can be obtained in SO(10) GUT model even with the simplest choice of yukawa coupling. Thus we can see the advantage of Generalized Froggatt-Nielsen mechanism over ordinary Froggatt-Nielsen mechanism in obtaining the standard model flavor structure in GUT models.

\section{\label{sec-3}Conclusions}

In this paper, we propose a Generalized Froggatt-Nielsen mechanism in which non-renormalizable operators involving a GUT group and $U(1)_H$ non-singlet Higgs field are introduced. Thus the GUT gauge symmetry breaking and the generation of hierarchical flavor hierarchy have a common origin in this mechanism.  In this Generalized Froggatt-Nielsen mechanism, we propose universality conditions for coefficients corresponding to different contractions in the group productions. We find that the predictions in Generalized Froggatt-Nielsen mechanism for SU(5) GUT is different to that of ordinary Froggatt-Nielsen mechanism. Such Generalized Froggatt-Nielsen mechanism can be used in GUT models when ordinary Froggatt-Nielsen mechanism is no longer available. We study the applications of Generalized Froggatt-Nielsen mechanism in SO(10) models. We find that realistic standard model mass hierarchy and mixings can be obtained both in SU(5) and SO(10) GUT models with such Generalized Froggatt-Nielsen mechanism.

It is also possible to abandon the universality conditions for coefficients corresponding to different contractions in the group productions. Then we can specify the desired representations from $\Phi_{r}^n$ and calculating the corresponding Clebsch-Gordon coefficients. This scenario is only natural in SUSY GUT models due to the non-renormalization theorem for superpotential\cite{Seiberg}. The effects of non-renormalizable superpotential involving high dimensional representation Higgs can be found in our previous works\cite{fei1,fei2,fei3}. We will discuss the scenario with non-universality conditions in our future works.

\begin{acknowledgments}
This research was supported by the Australian Research
Council under project DP0877916 and by the Natural Science Foundation
of China under project "The Low Energy Effects of Non-Renormalizable Terms in GUT Models".
\end{acknowledgments}

\end{document}